# Wave approach for the resonances of rectangular and triangular membranes

Joseph Dickey
3960 Birdsville Rd., Davidsonville, Maryland, 21035, USA
Joe@JoeDickey.com

**Abstract:**

This study develops a ray technique for determining the resonance frequencies of triangular membranes. The technique is demonstrated for homogenous rectangular and triangular membranes with fixed boundaries. Where possible, the results are compared with exact calculations. The membrane resonances are calculated using an equivalent string whose length is proportional to the *reciprocal* of the length of *closed* paths starting from an arbitrary point within the membrane. Closed paths are ray paths which arrive back at the starting point going in the same direction.

# Wave approach for the resonances of rectangular and triangular membranes

I. Introduction

This paper describes a wave, or ray, technique for determining the resonance frequencies of rectangular and triangular membranes. The technique probably has application to a larger class of irregular polygonal membranes, but this has not yet been demonstrated. The procedure is to start at an arbitrary interior point of the membrane and to generate a family of rays by incrementing the starting angle. The subset of these rays which form closed paths, i.e. paths which return to the origin going in the same direction, are identified. A set of one-dimensional systems, strings, is formed where the lengths of the strings are proportional to the reciprocal of the closed path lengths. The resonances of the membrane are then the resonances of the strings. Phase changes as the closed path impinges on boundaries must be incorporated in the string.

The present demonstration of the approach is restricted to 2-D planar homogenous isotropic membranes confined by polygonal boundaries with locally reacting impedances which do not vary along a given section of the polygon and where the angle of reflection equals the angle of incidence. Not all of these restrictions are necessary for the technique to work, but it is necessary that the boundary impedance be specified everywhere. Under these restrictions, the resonance behavior of three regular polygonal membranes with fixed edges; specifically, rectangles, and right isosceles and equilateral triangles are accurately determined. The more general case of the obtuse triangle is also studied and gives plausible resonance frequencies which converge to the regular cases as boundaries are made to approach these.

The motivation for my doing this was based on the intuition that a basic propagation (plane wave) around closed paths in an object should be related to the resonances of that object. Investigating this led to the discovery that closed path lengths in planar polygons are independent of starting location. This point deserves some scrutiny since it's not hard to construct polygons and/or their boundary conditions for which this is not true; but, it is true for the rectangles, triangles and associated boundary conditions used here. The association of the closed path lengths with a string length seemed logical, but the reciprocal nature of this was, at first, a surprise.

The regular polygons mentioned above were used as benchmarks in the present study because there are exact solutions available for membranes of these shapes with fixed boundaries. The squares and rectangles are treated in many text books. The right isosceles triangle was discussed by Rayleigh [1] and the equilateral triangle by Lamé [2].

There is one major problem which I have not yet resolved. I mentioned above that the lengths of the closed paths are related to the string lengths. More specifically, the string lengths are proportional to the *reciprocals* of the closed path lengths. The problem is that the constant of proportionality is not the same for all polygon shapes. For example; for rectangles the constant is 2*A where A is the area of the membrane, for the right isosceles triangle it is 4*A, and for the equilateral triangle it is 3*A. For the present approach to be useful, one needs to be able to derive the constant from the shape.

## II. Summary of the technique

1) Define the boundaries.
2) Select a single arbitrary starting point within the membrane.
3) Start a unit amplitude ray from this point at an angle which will be incremented to cover all angles.
4) Propagate this ray as a plane wave with propagation loss.
5) When the ray strikes a boundary, reflect it in accordance with the relative impedance of the membrane and the boundary, and continue propagating.
6) If the ray comes within a specified proximity of the starting point and within a specified proximity of the starting angle, the path is deemed *closed* and the path length is recorded, $(L_{cp})_N$. Stop the propagation and go back to step 3) with a new angle.
7) If closure fails, abandon this ray after the propagation loss has reduced the amplitude to some specified level. Go back to step 3) with a new angle.
8) Continue re-starting at step 3) until all angles are covered.
9) Form the lengths of the equivalent 1-D systems (strings), $L'_n$, from the set $L_{cp}$.
10) Remove duplicate lengths in $L'_n$.
11) The total resonances, i.e. the fundamental and all harmonics for each equivalent string, $L'_n$, is the complete set of resonances for the membrane.

Some discussion on the above.
1) There are lots of ways to define boundaries. Since calculations are here done on a computer (using MatLab®), I defined the boundaries as algebraic equations and "if" statements.
2) For the examples used here, i.e. rectangular and triangular boundaries and homogenous isotropic membranes, any starting point (within the membrane) will yield all possible closed paths.
3) Starting at angle = 0 makes sense and an increment on the order of $(2\pi/500)$ seemed to catch all the closed paths.
4) The plane wave propagation of an initially unit amplitude wave of angular frequency $\omega$ is of the form $\qquad \psi(x,t) \propto e^{-ikx+i\omega t}, \quad k = k_0(1-i\eta), \quad k_0 = \omega/c,$ (1)
where $\psi$ is any of the linear quantities, displacement, velocity, acceleration or pressure, $k$ is the wavenumber, $x$ is the distance propagated, $t$ is the time, $c$ is the wave's propagation velocity, and $\eta$ is the propagation loss factor. The step size as the wave propagates is important since it affects the accuracy of the reflection at the boundary. For all calculations here, the step size was 0.005.

5) The work here is only concerned with specular reflection where the angle of reflection equals the angle of incidence. The reflection coefficients are the plane wave coefficients determined by the boundary impedances relative to the membrane impedance. Without loss of generality, Dirchlet boundary conditions (boundary impedance = ∞ ) were used here.
6) There are some judicious choices here regarding when one decides that the path has *closed*. How close to the origin and how close to the starting angle? For the examples here, the membrane dimensions were order unity and the criteria for closeness to the wave starting position was 0.05 and the closeness to the starting angle was 0.05 radians.
7) If closure fails in a reasonable path length, there is need to terminate the ray. The total path length used here was generally 20. If the total path length is allowed to be too long, the ray may keep reverberating and eventually find itself within the closure criteria (depending on the closeness criteria above) and falsely get counted as a closed path. Even so, after a long propagation, the amplitude will be so weakened by propagation loss that any resonance, actual or false, would be very weak and should not be counted.
8) The angular increment used in searching for closed paths is also an important parameter that effects computation time and the certainty of finding closed paths. For regular polygons there is no need to cover $2\pi$ radians; a lesser quadrant will suffice.
9) The resonances associated with a particular closed path are calculated from the resonances of the equivalent string. The equivalent string length, $L'$, for a particular closed path length is related to the reciprocal of closed path length. The rationale for this is hinted at by noting that if a wave closes on itself to form a closed path, then, if allowed to continue, it will close on itself again, and again, etc. This unending series of increasing path lengths must correspond to the unending series of increasing frequencies (harmonics) of the equivalent string; but this is equivalent to the series of fundamentals of an unending series of *shorter* strings. Hence, the reciprocal relationship between $L_{cp}$ and $L'$. It is necessary to remove a) any paths that do not impinge all sides of the polygon and b) any paths which fold back on themselves.
10) Depending on the angular increments and the closeness criteria for closure, there may be multiple closed paths which are actually the same one. It's not really necessary to remove the duplicates, but not doing so makes each resonant frequency look like a closely spaced group.
11) Finally the combined resonances of all the equivalent strings is displayed by calculating

$$\psi_{res}(f) = \sum_N [1 - e^{-ikL'_N(1-i\eta)} e^{iM_N\pi}]^{-1}$$

the real of: $M_N$

$N^{th}$

This is the point at which the boundary conditions are handled. The accumulated phase change upon reflection must be accounted for. For example, for Dirichlet boundary conditions on all sides, and if there are $M_N$ reflections in the $N^{th}$ path, then the above equation becomes

$$\psi_{res}(f) = \sum_N [1 - e^{-ikL'_N(1-i\eta)} e^{iM_N\pi}]^{-1} \qquad (2)$$

The technique is illustrated by calculating the resonances of a rectangle, a right isosceles triangle, and an equilateral triangle; all of which can be solved exactly. Lastly, I calculate the resonances of an obtuse triangle, for which we have no analytical solution. For the examples considered, these shapes will be fixed on all sides although, as mentioned, the technique can be used for other boundary conditions; e.g. free, impedance, or combinations of these. The extension to inhomogeneous boundary conditions, e.g. boundary impedances which vary along their length, is not straightforward and not allowed in the formulation presented. I believe it is possible to do this, but I have not yet done so. The inhomogeneous boundary can probably be related to a limiting case of considering the side to consist of a number of segments, each with different but homogenous conditions. A problem arises since, in this case, a closed path may not impinge on all sides and therefore the missed segments will not influence the resonance frequencies; clearly they must. One could then choose a starting point such that the ray did impinge on a particular segment but this illuminates another problem; namely, that the resonance determination now depends on starting location. The many-sided polygon is similarly not trivial and is another area which I have not yet investigated.

I chose the triangle to illustrate the technique because it does not fit any of the coordinate systems in which the wave equation separates, and therefore we don't have exact solutions to give us the resonance frequencies. The two lone exceptions to this are the right isosceles and the equilateral triangles. The right isosceles, fixed along its perimeter can be considered as half a square and, as pointed out by Rayleigh [1], its modes must be a sub-set of the modes for the square; specifically, the modes which have nodes along the diagonal (i.e. the base of the triangle). The wave equation can be separated in the equilateral triangle; a result first pointed out by Lamé [2] and expanded upon for both fixed and free perimeters (Dirichlet and Neumann boundary conditions respectively) by McCartin [3].

## II. Determining resonant frequencies.

### A. The String

First consider a finite string driven harmonically at some arbitrary point along its length. A resonance frequency in the string can be viewed as a frequency for which, after one complete reverberation, the wave arrives back at the starting point going in the same direction. For some values of driving frequency, the eigenfrequencies, it will arrive with the same phase; thereby reinforcing the wave emanating from the drive. For the simple case where both ends are either free or fixed, these frequencies are;

$$f_n = nc/2L = nc/L_{cp}, \quad f = \omega/2\pi, \tag{3}$$

where *n* is a positive integer, c is the wave speed, *L* is the length of the string and the length of the closed path $L_{cp}$ is just 2*L*. This is clearly true for any starting point on the string.

## B. The Rectangle

When applying the above resonance criteria to a two-dimensional object, such as a rectangle, we would need to pick a starting point and a starting angle. As with the string, the starting point is arbitrary, but the angle is not. We would need to search for angles which would give closed paths; again, those unique paths wherein the wave passes through its point of origin going the same direction as when it started. This coincidence establishes a length for the closed path. As with the string, we postulate an harmonic drive at the starting point, and keep track of the phase of the wave as it propagates along this path including and phase changes as it reflects from the boundaries. If, when the wave closes on itself, this cumulative phase change equals an integer multiple of $2\pi$, we have a resonance. There will be a number of closed paths. The shorter paths are the stronger (more important) resonances because in the longer paths the wave will have lost more amplitude (because of propagation loss and less than unity reflections) and thereby provides less reinforcement of the original wave.

The classical analysis for the rectangular membrane with all sides either fixed or free is

$$f_{m,n} = \frac{c}{2}\left[\left(\frac{m}{L_x}\right)^2 + \left(\frac{n}{L_y}\right)^2\right]^{1/2}, \quad \begin{array}{l} m = 1,2,3 ..... \\ n = 1,2,3 ..... \end{array} \qquad (4)$$

The four shortest closed paths for a 1.37 aspect ratio rectangle, specifically with $(L_x : L_y) = $ (1:1.37), are shown in Fig. (1). For this simple geometry, the lengths of the closed paths are simply $(L_{cp})_{n,m} = 2\left[(nL_x)^2 + (mL_y)^2\right]^{1/2}$, and each of these lengths is associated with a resonant frequency given by $f_N = c/(L_{cp})_N$ where $N$ simply numbers the various $(n,m)$ lengths. These resonances must be the same as in Eq.(4) which leads to the definition

$$L'_N = \beta/(L_{cp})_N, \quad \beta = 2L_x L_y, \tag{5}$$

where $L'_N$ is the effective length of the $N$ one-dimensional systems (strings) represented by the closed paths. For rectangles, $\beta$ is also twice the area; this is not generally true for other polygons.

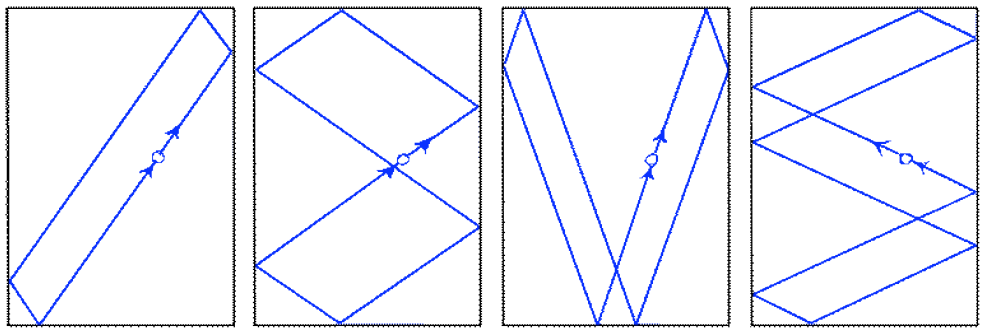

Fig. 1. The trajectories associated with the four shortest allowed closed paths for the rectangular membrane with fixed boundary and with aspect ratio = 1.37. The starting point is indicated by the circle.

The paths are one-dimensional and the propagation is plane wave. The resonance condition, as in the string, is when the phase matches the phase at the source and if this is taken to be zero, the resonances are given by the maxima of

$$\psi_{res}(f) = \sum_N [1 - e^{-i2kL'_N(1-i\eta)} e^{i\vartheta}]^{-1}, \quad k = \omega/c, \tag{6}$$

where $\eta$ is the propagation loss factor and $\vartheta$ is the accumulated phase change caused by reflections. For the rectangle, every closed path involves an even number of reflections and for Dirichlet boundary conditions, each reflection is $\pi$ so $e^{i\vartheta} = 1$. The "response" presented in the figures is the real part of $\psi_{res}$.

The $L'_N$ are derived from a more complete list of closed paths than shown in Table 1. This table gives the angles, measured from the horizontal, $L_{cp}$ and $L'_N$ for the 4 shortest path lengths. The membrane is fixed along its perimeter and the wave speed is taken to be 100 and $\eta = 0.01$.

| Angle (Radians) | Path Length, $L_{cp}$ | $L'_{cp} = 2(L_x L_y)/L_{cp}$ | Resonance associated with $L'_n$, (Hz) | Corrisponding (m,n) from Equation (4) |
|---|---|---|---|---|
| 0.9430 | 3.3667 | 0.8139 | 61.9 | (1,1) |
| 0.5970 | 4.800 | 0.5708 | 88.5 | (1,2) |
| 1.2190 | 5.800 | 0.4724 | 106 | (2,1) |
| 2.7140 | 6.5667 | 0.4173 | 120 | (1,3) |

Table 1. The starting angles, path lengths and resonant frequencies for the four shortest (and most important) closed trajectories for the rectangle.

It should be noted that the paths for which the starting angles are 0 or $\pi/2$ never impinge on some sides where boundary conditions are imposed. Such waves can not participate in satisfying boundary conditions on these sides and are thereby not allowed. This reasoning also applies to the classical mode analysis in Eq. (4) in which the lowest values for *m* and *n* are one, i.e. not zero. It should also be noted that these disallowed paths are the only ones that fold back on themselves and this will be the criteria for disallowing paths in other shapes.

The correspondence between the classical mode notation and the trajectory diagrams can be simply seen by noting that *(m+n)* is half the number of times the closed path trajectory impinges on the horizontal or vertical boundary, respectively.

Equation (6) is calculated and shown in Figure (2). The resonant frequencies given by Eg. (4) are marked and labeled by their *(m,n)* values and vertical lines. Note that the (2,2) mode in the Figure is too close to the (1,3) mode to be well resolved.

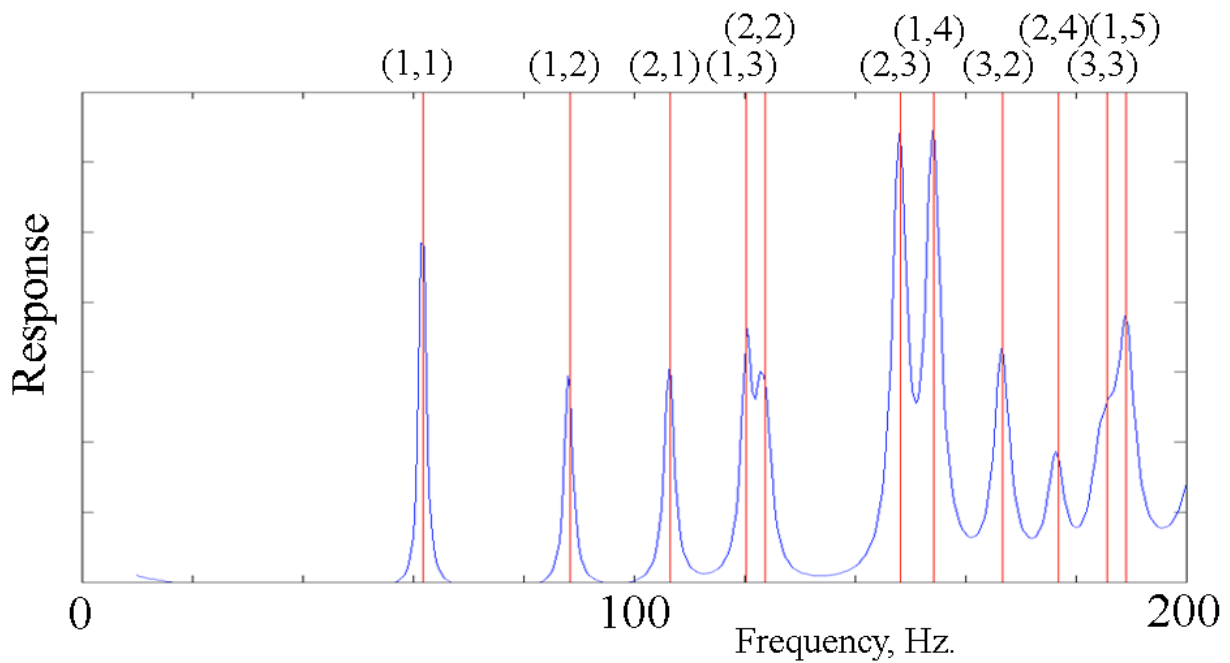

Figure 2. The resonant frequency response of a rectangular membrane with dimensions (1, 1.37), with fixed edges and a wave speed of 100. The curve is calculated by wave approach. The vertical lines identified by *(m,n)* at the top show the $f_{m,n}$ resonance frequencies calculated by Eq.(4).

## C. Triangles

### 1) The Right Isosceles Triangle

The straightforward application of the above procedure is now applied to a triangle. The triangular membrane considered here has unit sides in the *x* and *y* directions and is fixed on its perimeter.

It was noted by Rayleigh [1], that the modal frequencies of the fixed perimeter right isosceles triangle can be associated with a subset of the modes of the corresponding square. Specifically the square modes which have a nodal line along the diagonal. For the unit right isosceles triangle, these are,

$$f_{m,n} = \frac{c}{2}\sqrt{(m+n)^2 + n^2}, \quad m,n = 1,2,....; \tag{7a}$$

whereas the resonances for the unit square with fixed perimeter are given by,

$$f_{m,n} = \frac{c}{2}\sqrt{m^2 + n^2}, \quad m,n = 1,2,.... \tag{7b}$$

The closed path trajectories are determined in accordance with the method described above, and the first few are diagrammed in Fig. (3). The two shortest rays, represented by a) and b), fold back on themselves and are thereby disallowed. They would also give rise to resonant frequencies not allowed by Eq. (7a). It can be seen, by inspection, that the closed path length in a) of Fig. (3) is 2, i.e. twice the length of one of the equal legs, and that in b) it is $2\sqrt{2}$ times the leg length; from there on, it is not as obvious.

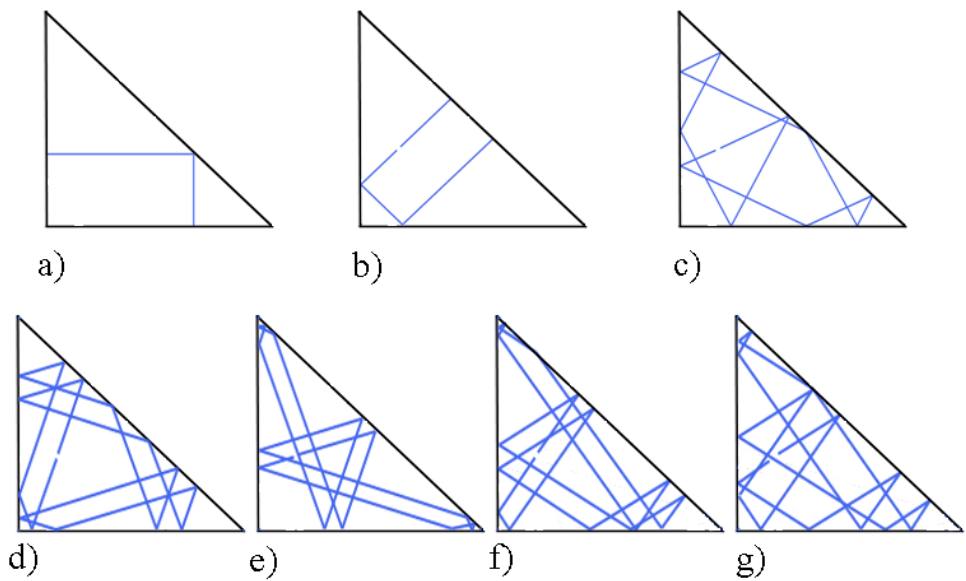

Figure 3. The first few closed path trajectories for the right isosceles triangle.

The factor $\beta$ in Eq. (5), which relates the closed path lengths with the effective length of the associated 1-D system is the same as for the rectangle; i.e. $\beta = 2L_x L_y$ where we still have $L_x = L_y = 1$; but now $\beta$ is four times the area of the triangle. It was confirmed that the starting angles which produce closed paths and their associated path lengths are independent of the starting location. This has to be, since otherwise the membrane's resonances would depend on where it was excited. This fact is borne out by examining paths f) and g). They are the same length and every leg of either path corresponds to a leg in the other which is propagating at the same angle. In fact, as the computer program sweeps through the angles, it will find a closed path at each angle present in the closed path. This leads to a computational advantage since every trajectory must at some point be traveling in, say, the first quadrant, and it is thereby only necessary to sweep that quadrant in searching for closed paths. This computational advantage does not hold for all polygons

Figure 4 shows the resonance response of the triangle calculated using the wave approach, specifically the resonance response function, Eq. (6). The solid vertical lines give the resonance frequencies determined by Rayleigh, Eq. (7a), and are labeled at the top with the indices for that equation. The correspondence between the wave calculation and the allowed resonances of Eq. (7a) is convincing and each peak in the wave calculation can be identified with one of the modes in the Rayleigh treatment. As mentioned, the Rayleigh resonances are a subset of the resonances of the enclosing square; i.e. there are square resonances which are *forbidden* in the triangle. The complete set of square resonances are calculated by Eq. (7b) and thereby include the *forbidden* ones which are shown by the vertical dashed lines in the figure. These forbidden modes are labeled with the indices as used in Eq. (7b) and identified in the figure in the square brackets near the bottom of the figure.

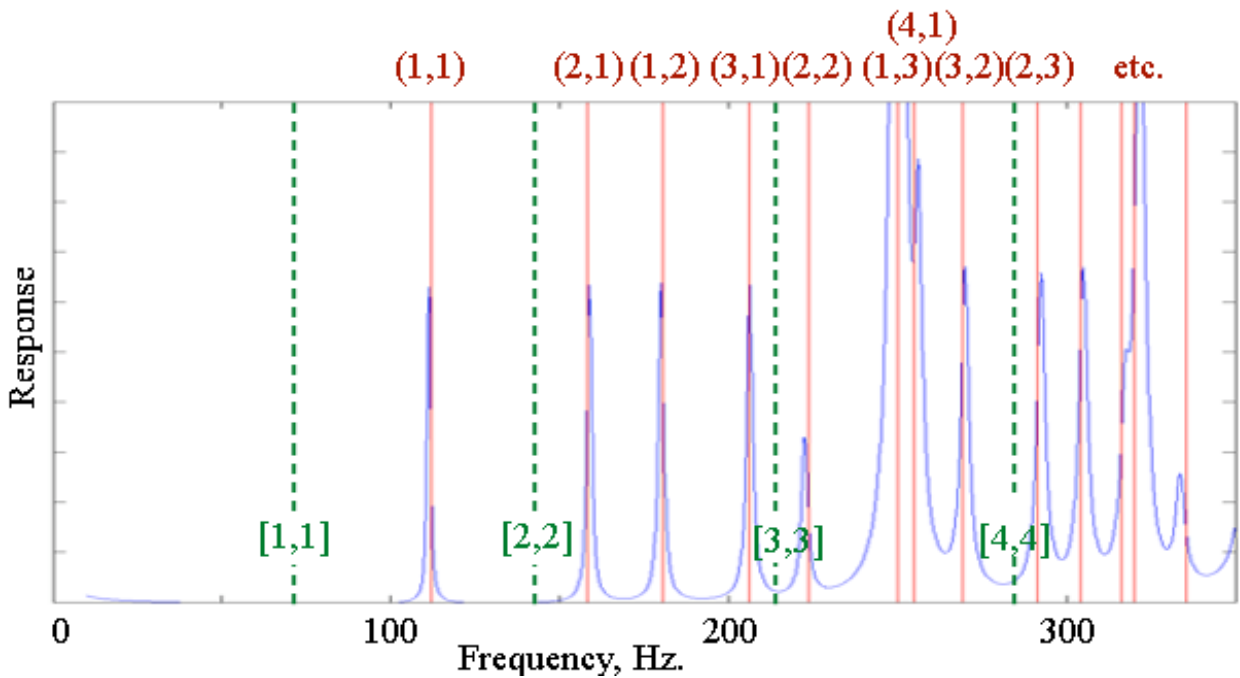

Figure 4. The resonant frequency response of a right isosceles triangular membrane with unit sides (and $\sqrt{2}$ hypotenuse) with fixed edges and a wave speed of 100. The curve is calculated by wave approach. The vertical solid lines, identified by *(m,n)* at the top, show the allowed $f_{m,n}$ resonance frequencies of a unit square calculated by Eq. (7a) The dashed lines show the *forbidden* resonances, $f_{m,n}$, identified near the bottom by [*m,n*], of the unit square calculated by Eq. (7b).

2) The Equilateral Triangle

The equilateral triangle can also be solved exactly. This was done by Lamé [2] who gives the resonance frequencies for the unit equilateral triangle as:

$$f_{m,n} = \frac{2c}{3}\sqrt{m^2 + n^2 + mn}, \quad m,n = 1,2,\dots . \tag{8}$$

The eight shortest closed paths for the equilateral triangle are ordered by increasing length and shown in Fig. (5). The correspondence between the closed path length, $(L_{cp})_N$, and the equivalent string length, $L'_N$, is:

$$L'_N = \beta/(L_{cp})_N, \quad \beta = 3\sqrt{3}/4 . \tag{9}$$

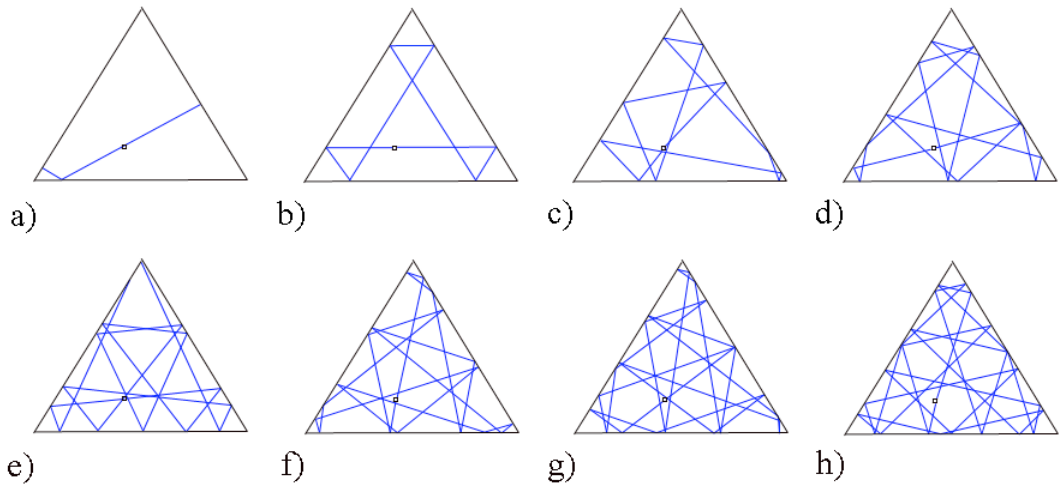

Figure 5. The first few closed path trajectories for the equilateral triangle.

The shortest path, a) in Fig. (5), folds back on itself and thereby not allowed. All other paths are allowed and the resonance response for the equilateral triangle with unit sides is calculated by Eq. (6) and shown in Fig. (6). The solid vertical lines give the resonance frequencies determined by Lamé and are labeled at the top with the indices for Eq.(8). The correspondence between the wave calculation and the allowed resonances of Eq. (8) is quite good and each peak in the wave calculation is identified with one of the modes in the Lamé treatment.

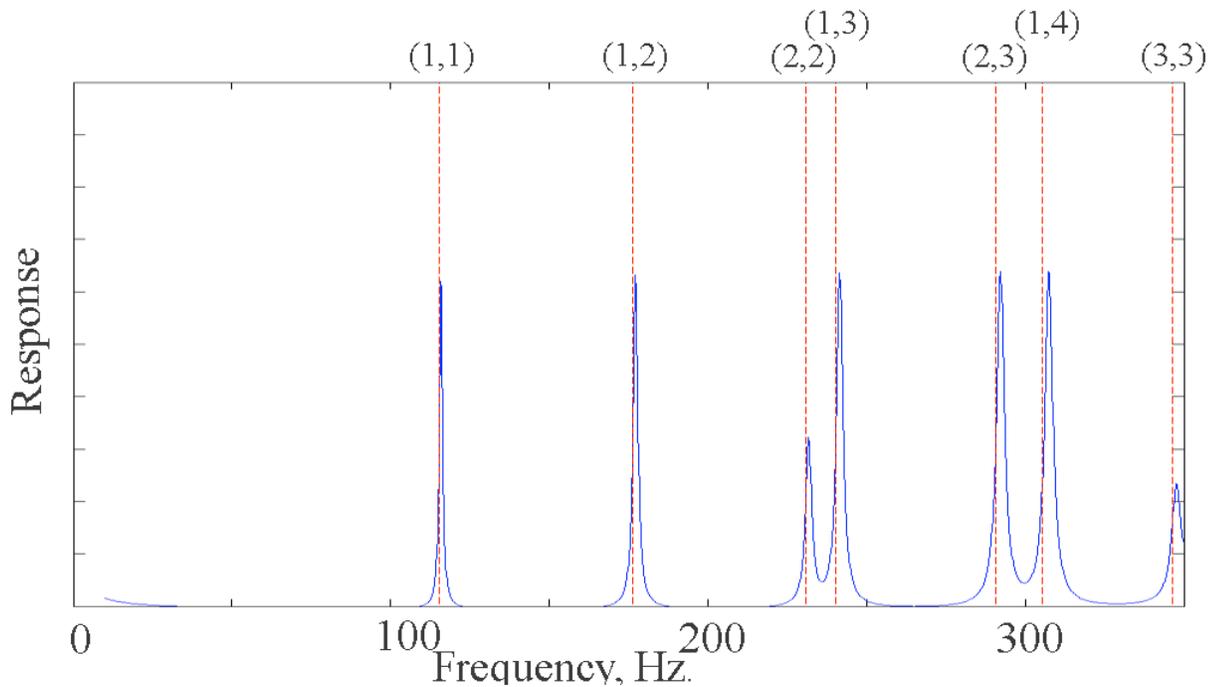

Figure 6. The resonant frequency response of an equilateral triangular membrane with unit sides and fixed edges and a wave speed of 100. The curve is calculated by the wave approach. The vertical lines identified by *(m,n)* at the top show the $f_{m,n}$ resonance frequencies calculated by Lamé's formula, Eq. (8).

2) The Obtuse Triangle

The technique is now applied to an obtuse triangle with arbitrary angles and a unit base; a shape for which there is no simple classical solution. The first few closed paths are shown in Fig. (7). Paths c) and d) are the same, though found from different starting angles, I confirmed that the resonant path lengths (and hence, the resonant frequencies) are independent of starting location. The angles which the sides make with the base in this example were 61 and 48 degrees.

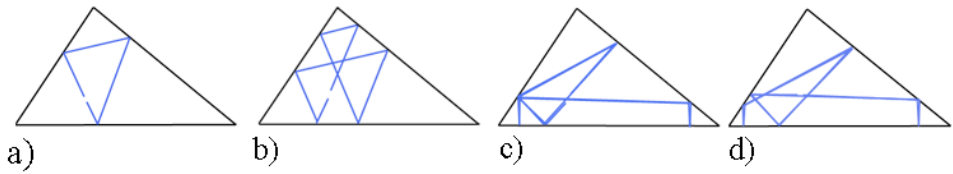

Figure 7. The first few closed path trajectories for the obtuse triangle.

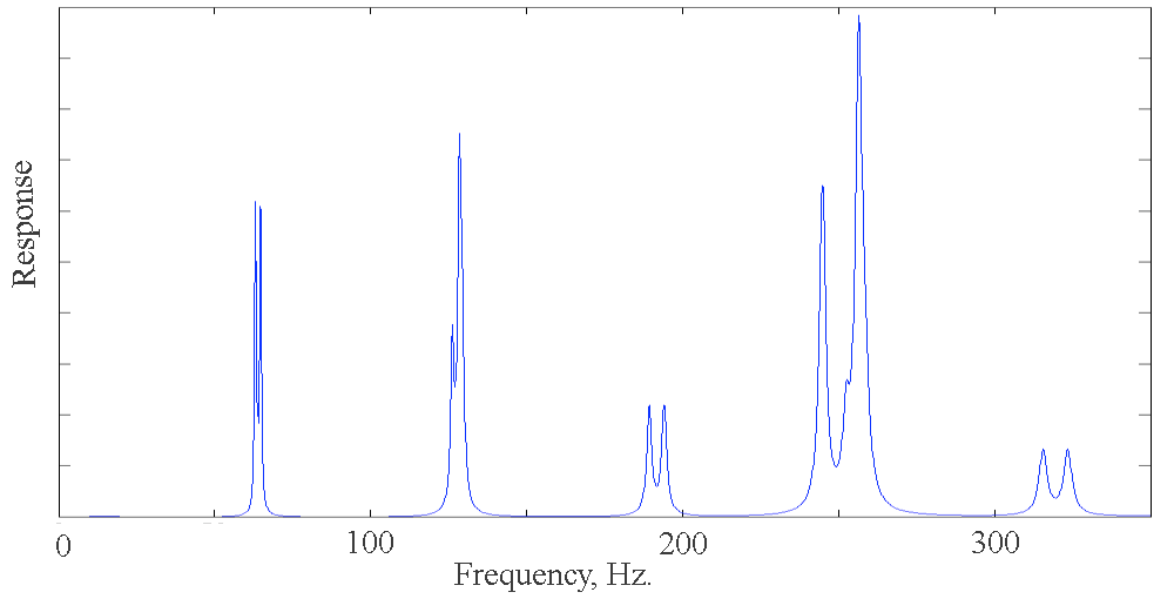

Figure 8. The calculated resonant frequency response of an obtuse triangular membrane with unit base, fixed edges and a wave speed of 100.

The splitting in the lowest frequency peaks results because the computer did not differentiate between two paths which were nearly the same path, but were started at different angles, and each met the closure criterion. This phenomena deserves a little more work.

## III. Conclusions

I have shown a way to calculate the resonant frequencies of rectangular and triangular membranes and demonstrated this by comparing the results with known results of exact calculations where they exist. I have extended this to the general obtuse triangle where exact solutions do not exist and the results look plausible. For the canonical shapes, rectangles, right isosceles and equilateral triangles, the comparison is remarkable; indicating that something is clearly correct.

The technique defines a relationship between the lengths of closed path ray trajectories in the polygon and the lengths of associated strings whose resonances are the resonances of the polygon. The relationship is that the associated string lengths are proportional to the reciprocal of the closed path lengths with the cumulative phase changes upon reflection at the boundaries accounted for. A serious shortcoming in this work is that I do not have a way to get this proportionality constant for the general polygon; just in cases where the lowest resonant frequency of the polygon is known. I welcome any insight readers might provide.